\documentclass[letter,
               biblatex,     
               keeplastbox,   
               hyphens,      
               ]{jacow}
\usepackage{pdfpages,multirow,ragged2e} %
\usepackage{caption,subcaption} %
\makeatletter%
	\ifboolexpr{bool{xetex}}
	 {\renewcommand{\Gin@extensions}{.pdf,%
	                    .png,.jpg,.bmp,.pict,.tif,.psd,.mac,.sga,.tga,.gif,%
	                    .eps,.ps,%
	                    }}{}
\makeatother

\ifboolexpr{bool{xetex} or bool{luatex}} 
 {}                                      
 {\usepackage[utf8]{inputenc}}           

\usepackage[USenglish]{babel}

\ifboolexpr{bool{jacowbiblatex}}%
 {%
  \addbibresource{llrf23.bib}
  \addbibresource{biblatex-examples.bib}
 }{}
\begin{document}

\title{LLRF system considerations for a compact, commercial C-band accelerator using the AMD Xilinx RF-SoC\thanks{This material is based upon work supported by the Defense Advanced Research Projects Agency under Contract Numbers 140D0423C0006 and 140D0423C0007. The views, opinions, and/or findings expressed are those of the author(s) and should not be interpreted as representing the official views or policies of the Department of Defense or the U.S. Government.}}

\author{J. Einstein-Curtis\thanks{joshec@radiasoft.net}, J. Edelen, B.
Gur, M. Henderson, G. Khalsa, M. Kilpatrick, \\ R. O’Rourke, RadiaSoft LLC, Boulder, USA \\
  R. Augustsson, A. Diego, A. Smirnov, S. Thielk, RadiaBeam Technologies, LLC \\
  B. Hong, Z. Li, C. Liu, J. Merrick, E. Nanni, L. Ruckman, S. Tantawi, \\ F. Zuo, SLAC National Accelerator Laboratory}

\maketitle

\begin{abstract}
  This work describes the LLRF and control system in use for a novel
  accelerator structure developed for a compact design operating in
  C-band developed by SLAC, with collaboration from RadiaBeam and
  RadiaSoft. This design is a pulsed RF/pulsed beam system that only
  provides minimal monitoring for control of each two-cavity pair.
  Available signals include only a forward and reflected signal for each
  pair; such a design requires careful consideration of calibration and
  power-on routines, as well an understanding of how to correct for
  disturbances caused by the entire RF signal chain, including a new SSA,
  klystron, and distribution system. An AMD Xilinx RF-SoC with a separate
  supervisory computer is the LLRF system core, with on-board
  pulse-to-pulse feedback corrections. This work presents the current
  status of the project, as well as obstacles and manufacturing plans
  from the viewpoint of developing for larger-volume manufacturing.
\end{abstract}

\section{INTRODUCTION}

Particle accelerators are increasingly seeing use in industry and
government for a broad range of uses, including: environmental
treatment, a variety of manufacturing, and sterilization. For many of
these use cases, a smaller, or portable, electron accelerator, with the
capability of operating on battery would be valuable. Here we present
the design of a low-level RF system for one of these proposed C-band
designs operating at 5.7 GHz, utilizing an AMD Xilinx
RF-SoC evaluation kit, the ZCU216\cite{ZynqUltraScaleRFSoC,ZynqUltraScaleRFSoCb}.

In addition to the use of a combined RF/signal processing system, the
design utilizes a small, bespoke klystron design along with
conventional copper accelerating cavities to generate a programmable
output energy between 2 and 20 MeV. The system has been designed to
minimize cabling and integration complexity, with only minimal RF
signals available for control and monitoring. For example,only a single
forward and reflected probe signal are for each cavity pair, requiring
back-calculation of the actual field for to meet the operational
requirements of the system.

In order to meet the needs of a battery-operated system, we want to
minimize the power used by each component in the LLRF system. Our
long-term goal is to minimize size, weight, power and cost (SWaP+C). The
RF system for each accelerating cavity pair includes a channel each of
RF drive power and reflected power, a solid state amplifier (SSA), and
novel klystron based on previous work at SLAC designing a
\emph{klystrino}\cite{scheitrumKlystrinoHighPower2000}.

\section{SYSTEM CONSIDERATIONS}

RadiaSoft is also responsible for developing a high-level controls
interface for the entire system. This is planned to run on a low-power
embedded device with a touchscreen, possibly on a remote, plug-in
terminal to ensure system safety. While EPICS is planned as the system
of choice for data transport and system monitoring, minimizing power
means using EPICS in a different environment from its normal use case
with always-on devices. Long-term plans consist of using standard
technologies that can be Ethernet-controlled (e.g. PoE) to allow for
minimizing the use of custom components. This includes minimizing the
non-recurring engineering cost and offloading risk by by utilizing
off-the-shelf RF-SoC devices that support robust power management
schemes to minimize engineering and design effort.

\subsection{Simulation}

\begin{figure}[!h]
        \centering
        \begin{subfigure}[h]{0.475\textwidth}
           \centering
           \includegraphics*[width=0.9\columnwidth]{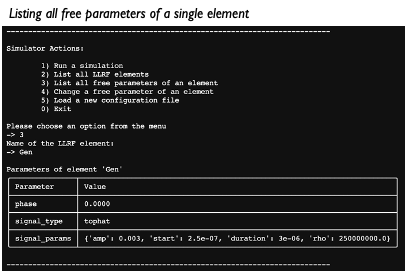}
           \caption{RF simulator high-level command-line interface.}
           \label{fig:simulator-menu}
        \end{subfigure}
        \begin{subfigure}[h]{0.475\textwidth}
           \centering
           \includegraphics*[width=.7\columnwidth]{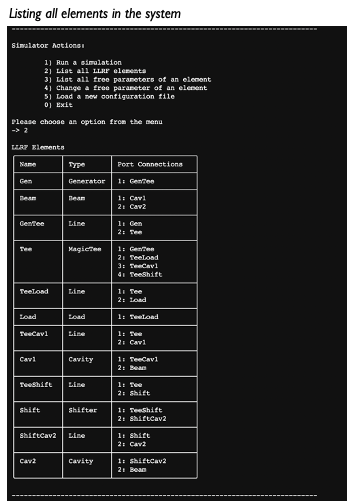}
           \caption{Example parameter dump from simulator command line.}
           \label{fig:simulator-parameters}
        \end{subfigure}
        \caption[ RF simulator control interface ]
        {\small RF simulator control interface. }
        \label{fig:simulator-usage}
\end{figure}

A full-featured, simplified-model RF simulator was developed to derive
control algorithms and characterize system performance. Controlling two
cavities requires a complete understanding of the signals being seen,
as well as being able to understand signal performance in the presence
of detuning or errors, see Fig.~\ref{fig:simulator-usage}.

A key feature of the simulator design was to allow for easy integration
into existing controls infrastructures to test algorithms. Key
measurements and settings are exposed as
EPICS\cite{EPICSExperimentalPhysics} process variables to allow for
easy code reuse and online simulation. This architecture allows for
easy testing of algorithms for signal correction. In particular, we
want to correct for both inter- and intra-pulse signal variation, as
well as system non-linearities, to be able to meet design requirements.

\begin{figure*}[!h]
        \centering
        \begin{subfigure}[h]{0.475\textwidth}
           \centering
           \includegraphics*[width=0.7\columnwidth]{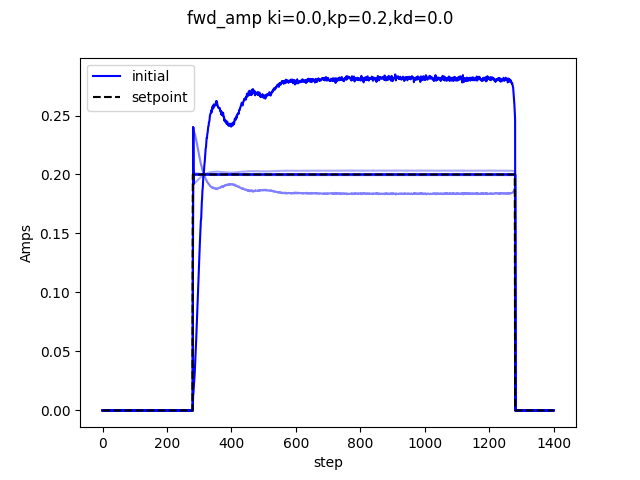}
           \caption{PID output signal waveform.}
           \label{fig:pid-over-time}
        \end{subfigure}
        \begin{subfigure}[h]{0.475\textwidth}
           \centering
           \includegraphics*[width=.7\columnwidth]{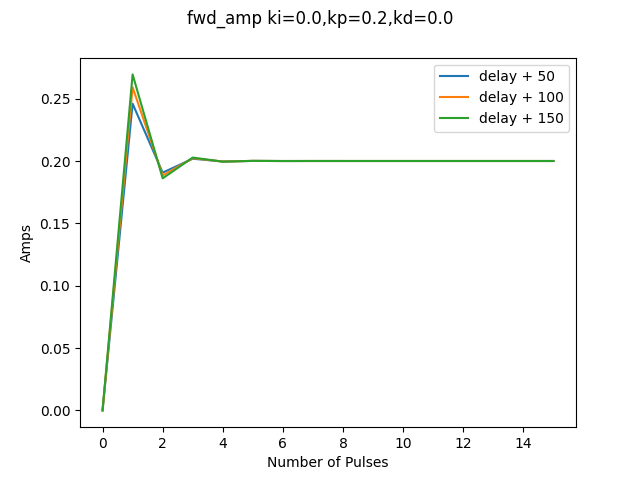}
           \caption{PID pulse-by-pulse optimization to setpoint.}
           \label{fig:pid-optimization}
        \end{subfigure}
        \caption[ Simulated pulse-by-pulse PID controller ]
        {\small Simulated pulse-by-pulse PID controller. }
        \label{fig:pid-pulse-by-pulse}
\end{figure*}

\begin{figure*}[!h]
        \centering
        \begin{subfigure}[h]{0.475\textwidth}
           \centering
           \includegraphics*[width=0.7\columnwidth]{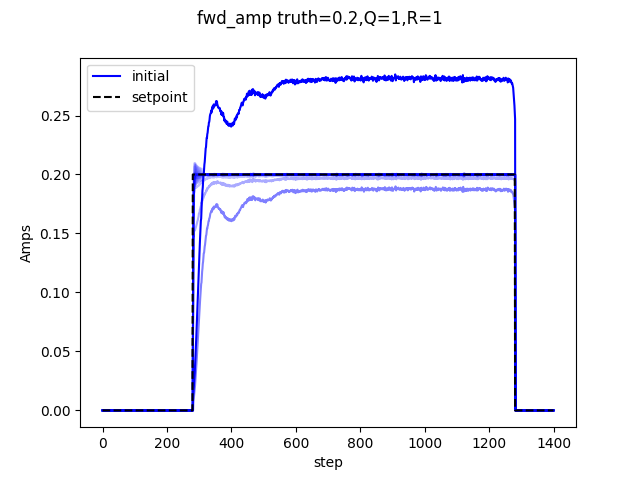}
           \caption{kalman output signal waveform.}
           \label{fig:kalman-over-time}
        \end{subfigure}
        \begin{subfigure}[h]{0.475\textwidth}
           \centering
           \includegraphics*[width=.7\columnwidth]{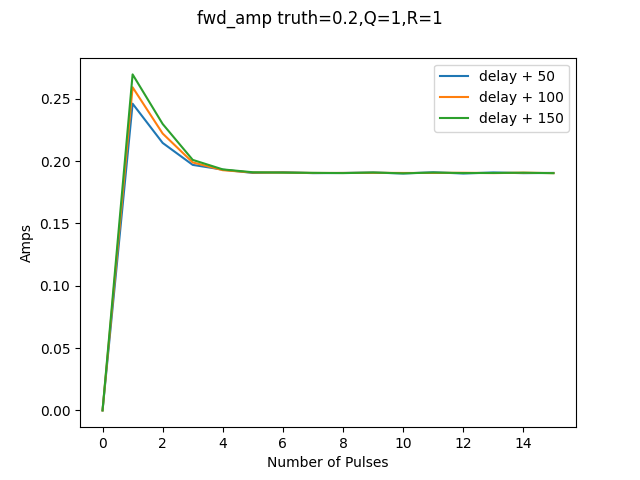}
           \caption{kalman pulse-by-pulse optimization to setpoint.}
           \label{fig:kalman-optimization}
        \end{subfigure}
        \caption[ Simulated pulse-by-pulse kalman controller ]
        {\small Simulated pulse-by-pulse kalman controller. }
        \label{fig:kalman-pulse-by-pulse}
\end{figure*}

For the prototype design, several control algorithms have been tested
using this simulator, including PID and Kalman methods. The
system design was loaded into the simulator with expected component
values and cavity tuning and several optimization routines were run to
test pulse-by-pulse flattening and amplifier linearization. PID control
can be seen in Fig.~\ref{fig:pid-pulse-by-pulse}, while the use of a
kalman-filer-based controller can be seen in
Fig.~\ref{fig:kalman-pulse-by-pulse}.

\subsection{SWaP+C}

Size, weight, and power, are the key drivers of portable systems.
Meeting these needs, while minimizing cost, is particularly difficult
on small-scale research and industrial systems.

While overall power budgets are still being determined, minimizing the
power used by the LLRF system itself is key. The RF-SoC device uses the
majority of the LLRF power budget, as unlike in systems with a single
RF source, particular consideration needs to be made of the power per
cavity. In addition, the plan is to run the system in pulsed operation,
allowing us to calculate both static and dynamic power, with static
power for each component of the LLRF control system for both sleep and
operational modes, as well as a dynamic power per pulse. Power
estimates for the implemented RF-SoC, 8-channel design can be seen in
Table~\ref{tab:rfsoc-power}.

\begin{table}[!hbt]
   \centering
   \caption{RF-SoC Power Estimation (14.454 W total)}
   \begin{tabular}{lcc}
       \toprule
       \textbf{Subsystem} & \textbf{Power (W)} \\
       \midrule{}
          \textbf{Dynamic} & \\
          Clocks & 0.250 \\
          Signals & 0.187 \\
          Logic & 0.178 \\
          BRAM & 0.239 \\
          URAM & 0.081 \\
          RFAMS & 9.660 \\
          DSP & 0.009 \\
          PLL & 0.172 \\
          I/O & 0.008 \\
          I/O & 0.008 \\
          SYSMON & 0.005 \\
          PS & 2.268 \\
          \textbf{Static} & \\
          PL & 1.289 \\
          PS & 0.108 \\
       \bottomrule
   \end{tabular}
   \label{tab:rfsoc-power}
\end{table}

Proper power management enables significant power savings in the LLRF
system across all its operational modes (e.g. sleep, running).
Engineering proper wake-up and sleep procedures significantly increases
the complexity of the software and systems architecture as many more
edge cases need to be considered in the design. For example, the RF
subsystem of the RF-SoC itself requires consideration of turn-on and
calibration times, with an at least 35 ms budget required to run the
turn-on state machine for an RF-SoC
\emph{tile}\cite{ZynqUltraScaleRFSoCa}. When operating the system at
higher repetition rates, this limits the ability to save static power
between pulses.

Additionally, proper RF component and frequency plan selection is
necessary to minimize wasted power. Spurs from digitization and losses
in baluns and other components require proper selection in the design
phase to avoid decreasing system efficiency.

\subsection{High-level control}

High-level control is a critical part of SWaP+C in such a system, with
the high-level communication and control interface having a significant
effect on the static power of the system. Significant advances in
Ethernet technology, along with the possibility for remote
power-over-ethernet (PoE), per-port management, open up a number of
opportunities for minimizing power on a per-device basis in this
system.

Unfortunately, no consumer devices utilizing the AMD Xilinx RF-SoC are
currently available that utilize PoE, requiring either a custom design
or consideration of additional power management schemes with a managed,
central power controller. Proper programming in a low-power, high-level
controller will help minimize SWaP through intelligent device
management. This is particularly important when considering the static
power required by support systems such as vacuum, environmental, and
safety systems.

Due to such system-level management needs, and its tight coupling with
the LLRF system, an NXP i.MX 8 device was selected for initial
prototyping, These devices are well-supported in both Linux and a
number of real-time operating systems (RTOS), allowing for a relatively
easy, and portable, iterative design process.

\section{Testing Results}

\begin{figure*}[!h]
        \centering
        \begin{subfigure}[h]{0.475\textwidth}
           \centering
           \includegraphics*[width=0.7\columnwidth,]{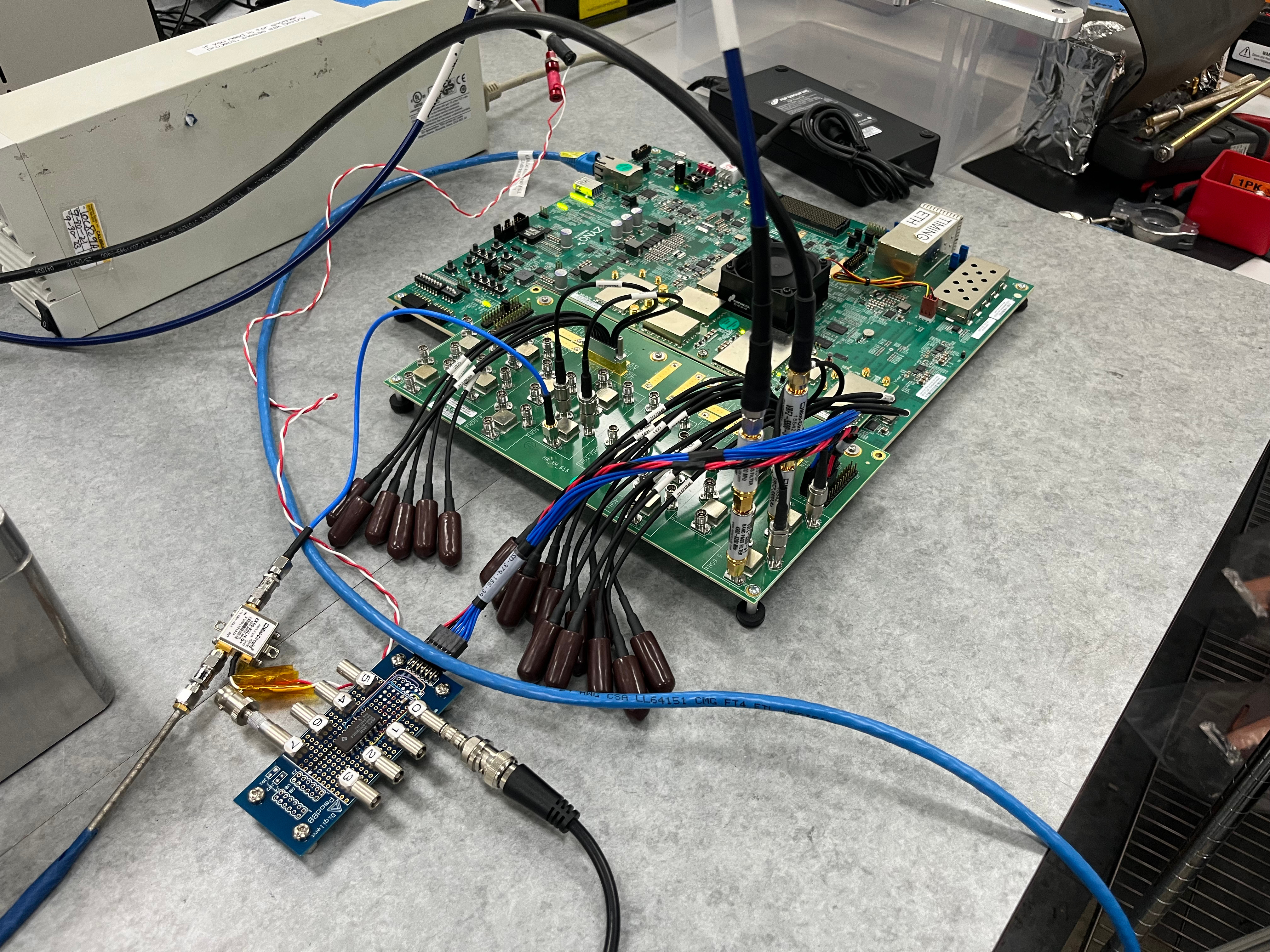}
           \caption{AMD Xilinx ZCU216 with trigger breakout board.}
           \label{fig:rfsoc-lab-device}
        \end{subfigure}
        \begin{subfigure}[h]{0.475\textwidth}
           \centering
           \includegraphics*[width=.7\columnwidth]{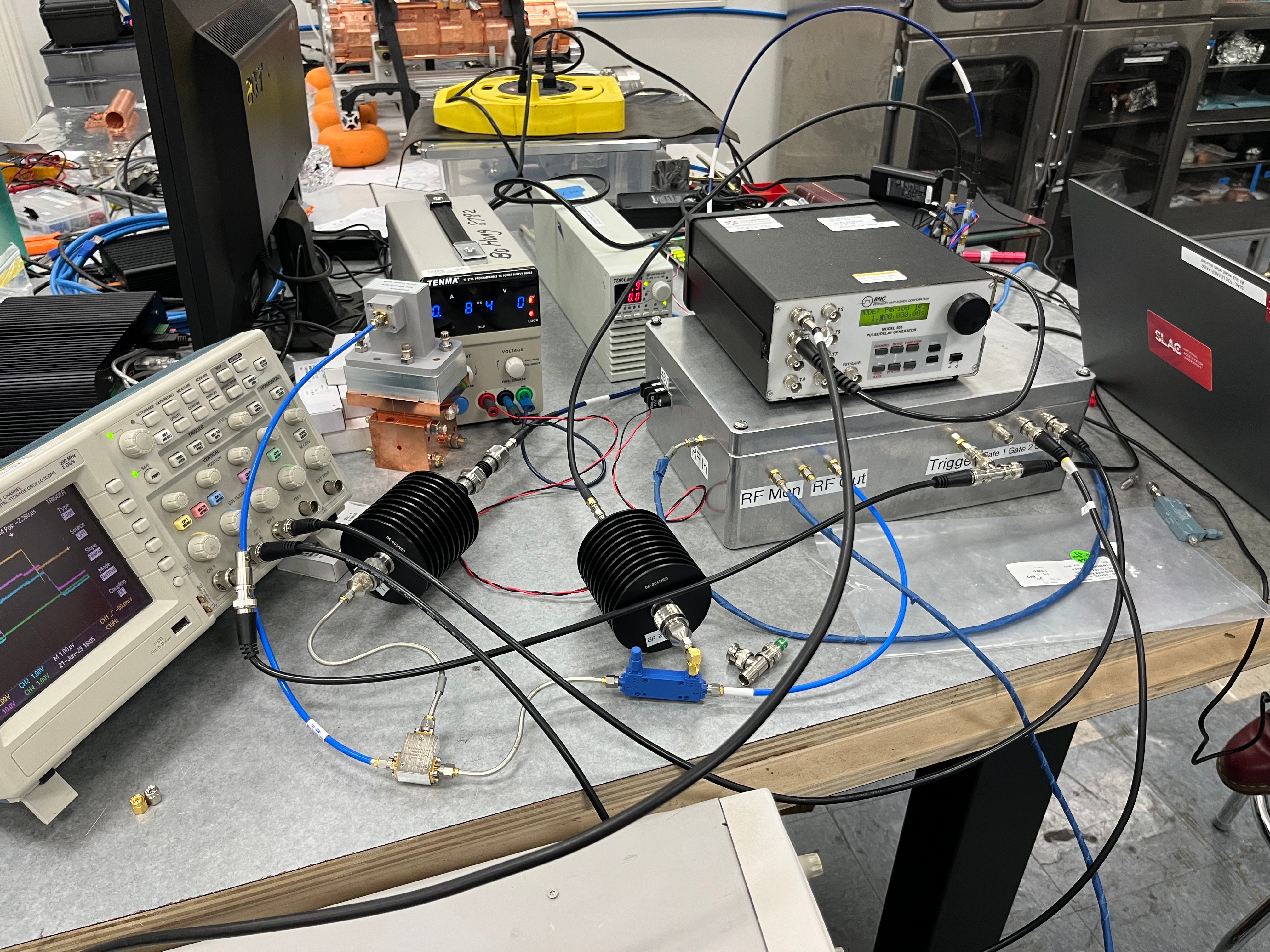}
           \caption{Test cavity and pulsed signal monitoring at RadiaBeam.}
           \label{fig:rfsoc-lab-2}
        \end{subfigure}
        \caption[ RFSoC cavity testing ]
        {\small Using the ZCU216 RF-SoC to drive a test cavity. }
        \label{fig:rfsoc-lab-testing}
\end{figure*}

\begin{figure*}[!h]
        \centering
        \begin{subfigure}[h]{0.475\textwidth}
           \centering
           \includegraphics*[width=0.9\columnwidth,clip,trim=4.5in 0 1.1in 5.7in]{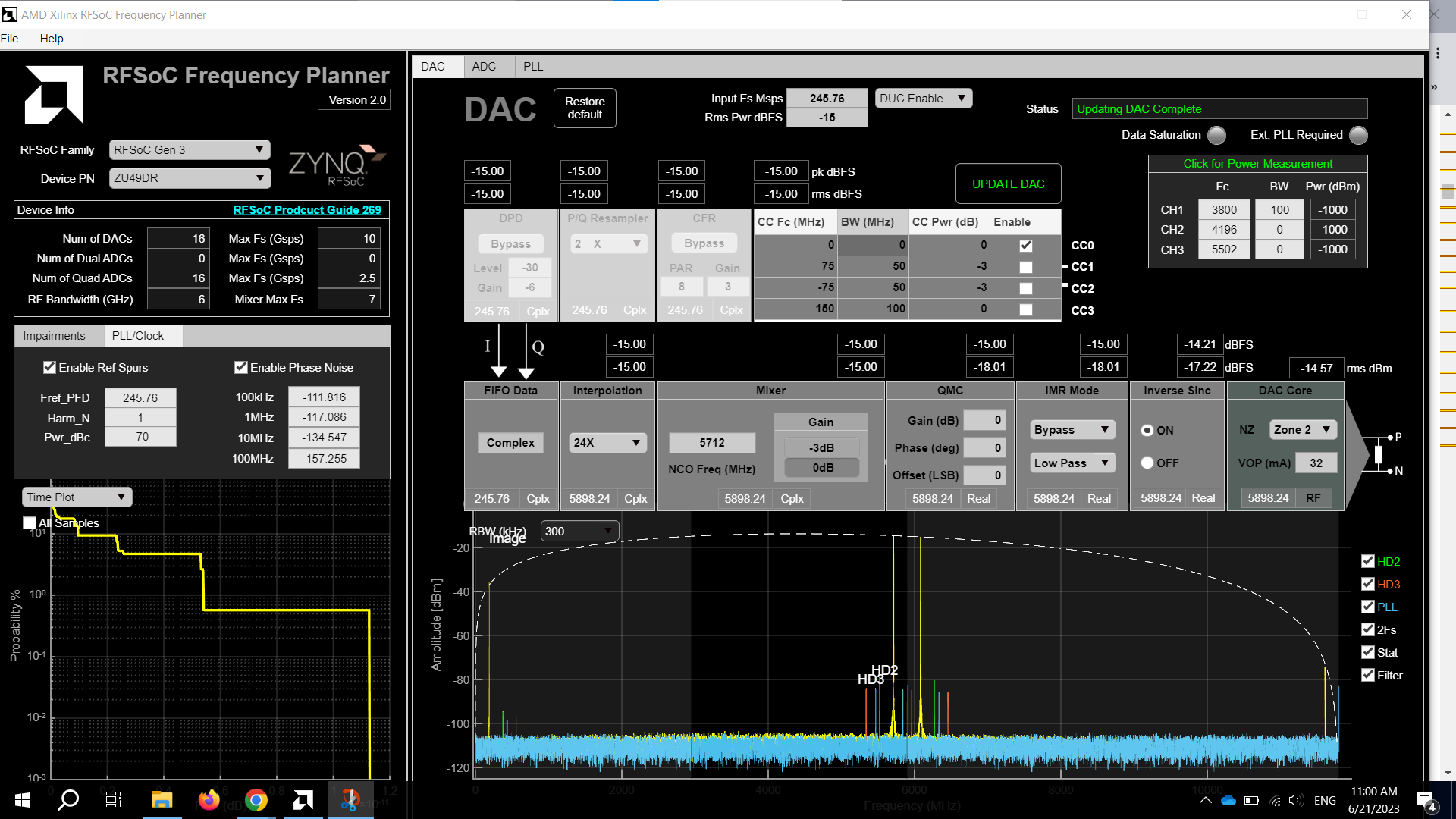}
           \caption{Simulated DAC output spectrum of ZCU216 using
           vendor-provided simulation tool.}
           \label{fig:rfsoc_simulated_dac_output}
        \end{subfigure}
        \begin{subfigure}[h]{0.475\textwidth}
           \centering
           \includegraphics*[width=.7\columnwidth]{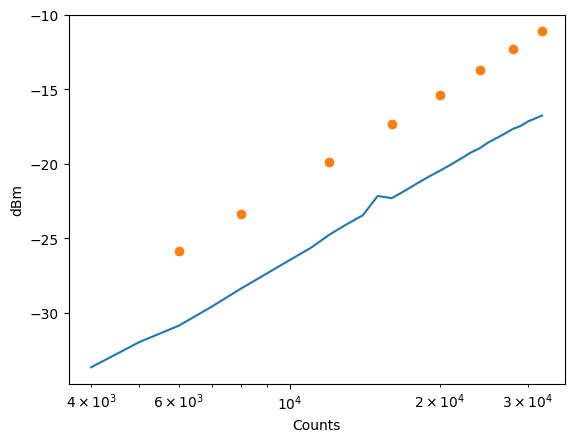}
           \caption{Measured RF power at frequency of interest using
           FFT (line) and using a power meter orange(dots) for ZCU216 at a given DAC output level in counts.}
           \label{fig:rfsoc-rf-power}
        \end{subfigure}
        \caption[ ZCU216 Output ]
        {\small AMD Xilinx ZCU216 output power consideration. }
        \label{fig:zcu216-rf-spectrums}
\end{figure*}

A series of tests on the ZCU216 and combined LLRF+SSA system was
performed at RadiaBeam, see Fig.~\ref{fig:rfsoc-lab-testing}. These
tests confirmed the output spectrum that was simulated using AMD Xilinx
tools, Fig.~\ref{fig:rfsoc_simulated_dac_output}. The prototype design
uses settings such that its clocks and sampling frequencies drive its
output in the second Nyquist zone, leading to a mirror of the desired
frequency in the third Nyquist zone of generally equal output power.

Additional measurements were taken to confirm these expectations, as
can be seen in the difference between power measured at the frequency
of interest using an FFT as opposed to those from a total power meter,
see Fig.~\ref{fig:rfsoc-rf-power}.

\begin{figure*}[!h]
        \centering
        \begin{subfigure}[h]{0.475\textwidth}
           \centering
           \includegraphics*[width=.7\columnwidth]{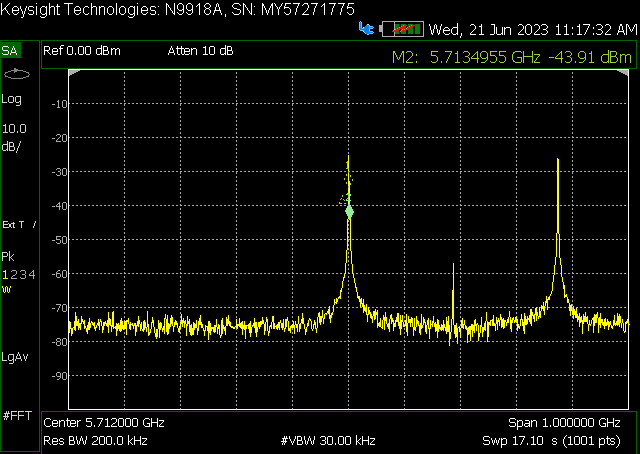}
           \caption{Output spectrum of the ZCU216 LLRF system.}
           \label{fig:zcu216-output-spectrum}
        \end{subfigure}
        \begin{subfigure}[h]{0.475\textwidth}
           \centering
           \includegraphics*[width=.7\columnwidth]{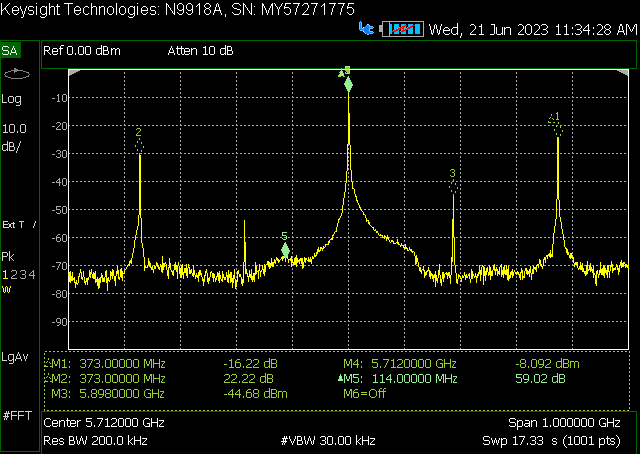}
           \caption{Output spectrum of SSA driven by the ZCU216 LLRF system.}
           \label{fig:rfsocssa-output-spectrum}
        \end{subfigure}
        \caption[ RF spectrum calculations]
        {\small RF spectrum calculations and measurements}
        \label{fig:measured-rf-spectrums}
\end{figure*}

Further testing confirmed this expectation when measuring spectrums at
the output of the LLRF system and at the output of the SSA, see
Fig.~\ref{fig:measured-rf-spectrums}. The two peaks represent the
frequency of interest and its image across the zone boundary,
with a small peak seen at an offset equal to the sampling frequency of
the output.

\section{CONCLUSION}

Here we present the design philosophy for an LLRF system designed for
portable applications using an AMD Xilinx RF-SoC-based LLRF system, as
well as the development of an associated, lightweight simulation
toolkit. An RF simulator has been used to develop several control
algorithms to correct for systematic and design variation. A prototype
LLRF system using evaluation hardware has been tested, with design
considerations for SWaP+C and RF performance reported here.



%
%
\ifboolexpr{bool{jacowbiblatex}}%
	{\printbibliography}%
	{%

	
} 

%
%


\end{document}